\documentclass{Interspeech}

\interspeechcameraready

\title{Voice-ENHANCE: Speech Restoration using a Diffusion-based Voice Conversion Framework}

\author[affiliation={1}]{Kyungguen}{Byun}
\author[affiliation={1}]{Jason}{Filos}
\author[affiliation={1}]{Erik}{Visser}
\author[affiliation={1}]{Sunkuk}{Moon}

\affiliation{}{Qualcomm Technologies Inc.}{USA}
\email{kyunggue@qti.qualcomm.com, jfilos@qti.qualcomm.com, evisser@qti.qualcomm.com, sunkukm@qti.qualcomm.com}
\keywords{Noise suppression, Speech restoration, Voice conversion, Diffusion models}

\usepackage{comment}

\begin{document}

\maketitle

\begin{abstract}
We propose a speech enhancement system that combines speaker-agnostic speech restoration with voice conversion (VC) to obtain a studio-level quality speech signal. While voice conversion models are typically used to change speaker characteristics, they can also serve as a means of speech restoration when the target speaker is the same as the source speaker. However, since VC models are vulnerable to noisy conditions, we have included a generative speech restoration (GSR) model at the front end of our proposed system. The GSR model performs noise suppression and restores speech damage incurred during that process without knowledge about the target speaker. The VC stage then uses guidance from clean speaker embeddings to further restore the output speech. By employing this two-stage approach, we have achieved speech quality objective metric scores comparable to state-of-the-art (SOTA) methods across multiple datasets.
\end{abstract}

\section{Introduction}
Speech enhancement (SE), also known as noise suppression (NS), is crucial in real-world applications such as telecommunications, voice assistants, and hearing aids. One notable limitation of SE systems is their performance in challenging signal-to-noise ratio (SNR) scenarios, where the enhanced speech often ends up being muffled or choppy \cite{wang2018supervised}. In particular, masking-based methods are prone to suppress target speech when the noise is speech-like or when the noise level dominates. Besides additive noise, speech can be impacted by a variety of simultaneously occurring acoustic distortions that can be highly non-linear in nature, such as reverberation, band-limitation, clipping, dropped packets, and codec artifacts. 

Speech restoration is an alternative but related approach to SE, where the target is not only to remove noise and undo acoustic distortions, but also to regenerate and restore speech to its full-band form and thus obtain studio-like quality \cite{liu2022voicefixer}\cite{zhang2021restoring}\cite{serra2022universal}\cite{babaev2024finally}\cite{abdulatif2022cmgan}. In situations where the original speech is either completely lost or severely corrupted, such mapping-based speech restoration may provide a better solution than masking-based methods. This is because in low SNR scenarios there is an inherent ambiguity that prevents the clean speech content from being restored clearly \cite{babaev2024finally}. Generative models, such as generative adversarial networks (GANs) \cite{goodfellow2020generative} or diffusion probabilistic models \cite{song2020score}, are typically used for this purpose.

Voice conversion (VC) is a promising avenue for speech restoration, aiming to transform speech from a source speaker to a target speaker while preserving linguistic content. Recent VC models \cite{wang2021vqmivc}\cite{freevc}\cite{kaneko2020cyclegan}\cite{diffvc} extract speaker-independent content features, such as phonetic posteriorgrams, from the input speech. Generative models then use these features to produce target speech with speaker-specific features, such as speaker embeddings. The training process for VC is similar to that of speech restoration, as it involves extracting speaker-specific features from the source speech and using a generative step to restore the original speech state under appropriate guidance.

Typically trained on clean data, VC models can generate high-quality outputs. However, they are susceptible to corruption when the input speech is noisy. This vulnerability necessitates the use of a multi-stage enhancement framework. For instance, integrating a noise suppression and speech restoration model at the front end can effectively handle noisy conditions before VC is applied. Such two-stage approaches have been validated in various studies, demonstrating improved performance in noisy environments \cite{strake2020speech}\cite{grais2017two}\cite{zhao2018convolutional}\cite{strake2019separated}.

\begin{figure*}[!t]
\includegraphics[width=\textwidth]{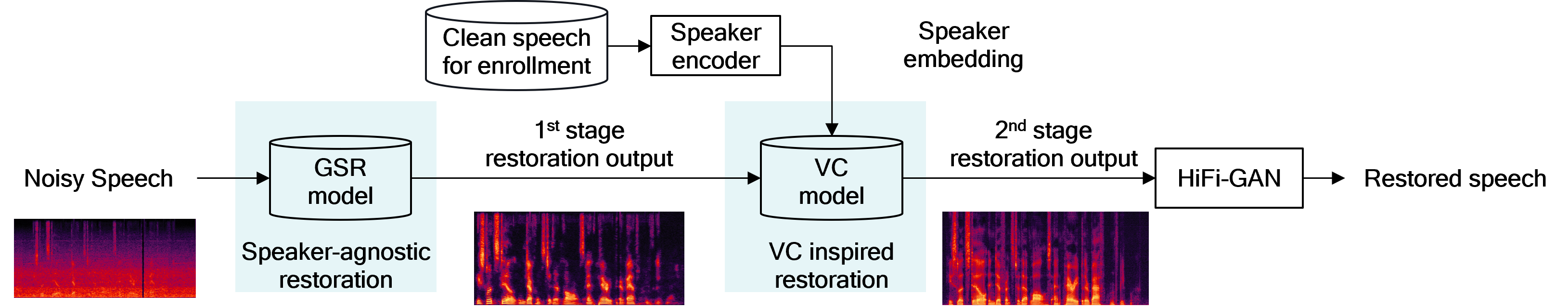}
\centering
\caption{\label{first} Overview of Voice-Enhance framework, noisy mixture is given to speaker-agnostic Generative speech restoration (GSR) module, then VC-inspired generative model generates high quality output guided by speaker embedding from the clean speech enrollment.} 
\vspace{-15pt}
\end{figure*}

In this paper, we propose an integrated framework that combines speech enhancement and restoration, and uses VC to obtain studio-quality speech. Our initial NS and restoration stage is based on a model called Generative Speech Restoration (GSR) which is derived from VoiceFixer \cite{liu2022voicefixer}. To further improve speech quality, a Diff-VC-inspired \cite{diffvc} VC model is employed, where the generative model is conditioned with a clean speech speaker embedding from the same speaker as the speech to be denoised. We assume short, uncorrelated segments of clean speech are obtained beforehand. The generative model was enhanced by replacing the content encoder with the vector-quantized (VQ) output of the HuBERT encoder for better linguistic feature extraction using self-supervised learning (SSL) features.
 
The primary objective of our research is to develop a robust speech restoration method that delivers studio-level speech quality while maintaining a reasonable model size. Our contributions include the novel combination of speaker-agnostic speech restoration (which itself outperforms the baseline in \cite{liu2022voicefixer}) and a diffusion-based voice conversion that uses SSL and speaker embeddings. The integration of GSR at the front end of our system addresses the vulnerability of VC models operating under noisy conditions. By employing this multi-stage approach, we achieve objective speech quality scores comparable to state-of-the-art methods across multiple datasets, highlighting the effectiveness of our system.

\vspace{-5pt}
\section{Background}
Speech restoration is a challenging task that aims to recover speech components from various degradation factors. We can define the degraded speech signal $x$ as $x = f(s) + n \in \mathbb{R}^{T}$, where $f$ represents various degradation factors, and $s$ and $n$ are the original speech and noise signals, respectively. It is common in speech enhancement to predict the speech directly from input $x$ via time-domain methods, or mask-based frequency-domain methods. The goal, however in speech restoration is to find a mapping function $g : \mathbb{R}^{T} \rightarrow \mathbb{R}^{T}$ that transforms the degraded speech signal $x$ into a high quality restored speech signal $\hat{s}$. The mapping function $g(\cdot)$ can also be seen as a generator and the problem decomposed into two-stages where the distorted speech $x$ is first mapped into a representation $z$, i.e. $f:x \rightarrow z$ and then a generator synthesizes $z$ to the restored speech $\hat{s}$, i.e. $g:z \rightarrow \hat{s}$. 

Diffusion models have shown extraordinary performance in generative tasks and are widely utilized in speech-related applications \cite{borsos2023audiolm}\cite{serra2022universal}\cite{gradtts}. The diffusion process, a stochastic method frequently used in generative modeling, consists of two primary stages: the forward process and the backward process \cite{song2020score}. In the forward process, noise is incrementally added to the data, transforming it into a noise distribution. The backward process aims to reconstruct the original data from the noisy data by reversing the forward process. During this reverse process, the model is trained to learn a scoring function that represents the gradient of the log probability density \cite{song2020score}. This framework allows for the generation of high-fidelity data by iteratively denoising the noisy input, making it a powerful tool in various applications.

HuBERT (Hidden-Unit BERT) \cite{hubert} is a self-supervised speech representation learning model designed to extract informative features without relying on a lexicon during pre-training. The HuBERT model is trained with a BERT-like \cite{devlin2018bert} prediction loss, focusing on masked regions to learn a combined acoustic and language model. Additionally, HuBERT employs an offline clustering step to generate discrete units, allowing it to transform continuous speech inputs into discrete hidden units and learn a robust, text-related speech representation.

\vspace{-5pt}
\section{Proposed method}
An overview of our proposed system is shown in Fig. 1. The noisy speech is first processed by the speaker-agnostic GSR model. In this work, we consider $z$ as the Mel-spectrogram of $x$ as our intermediate representation and use HiFi-GAN \cite{hifigan} as the generator. Then, a VC model generates restored speech by conditioning the diffusion process with a clean speech target speaker embedding. The clean speech used to extract the speaker embedding does not need to be content-aligned with the input. In this way, we further refine the restored speech by utilizing the SSL representation and speaker embedding as conditioning vectors for the speech restoration network $g$, such that $\hat{s} = g(x \mid h(x), m(x))$, where $h$, and $m$ represent the SSL network and speaker encoder network, respectively.

\vspace{-3pt}
\subsection{Generative Speech Restoration (GSR)}
Generative Speech Restoration (GSR) is a speaker-agnostic speech restoration network based on the ResU-Net analysis network proposed in \cite{liu2022voicefixer}. It is a comprehensive framework designed for high-fidelity speech restoration, capable of addressing multiple types of distortions such as noise, reverberation, limited bandwidth, clipping, packet loss, and codec artifacts. Unlike prior methods that focus on single-task speech restoration (e.g., only denoising or dereverberation), GSR aims to handle multiple distortions simultaneously, making it a more versatile solution for real-world applications.

The model consists of an analysis stage and a synthesis stage. In the analysis stage, GSR employs a ResU-Net architecture to predict enhanced intermediate-level speech features from the degraded input speech. In the synthesis stage, HiFi-GAN is used as the neural vocoder to generate the final high-fidelity waveform.

Unlike most masking-based speech enhancement and restoration methods, GSR also supports bandwidth extension and in-painting. Therefore, the restoration process involves an element-wise addition instead of element-wise multiplication \cite{abdulatif2022cmgan}. We modify the objective in \cite{liu2022voicefixer} such that $\hat{\mathbf{S}}_{mel} = f_{mel}(\mathbf{X}_{mel}; \alpha) + (\mathbf{X}_{mel} + \epsilon)$ where the mapping function $f_{mel}(\cdot, \alpha)$ is the ResU-Net mel-spectrogram restoration network parameterized by $\alpha$, and the output of $f_{mel}$ is added to $\mathbf{X}_{mel}$ to obtain the restored spectrogram, which is then used downstream. Unlike the denoising task of masking-based approaches, the network is not learning mask activations between 0 and 1 to suppress noise and preserve speech, but rather activations that can complete missing time-frequency (TF) bins \cite{abdulatif2022cmgan}.
We train GSR using the Generative Adversarial Network (GAN), Feature Matching (FM), and Mel-spectrogram losses from HiFi-GAN \cite{hifigan}.

\vspace{-3pt}
\subsection{VC-Inspired Speech Restoration}
The VC-based speech restoration, as shown in Fig. 2, is based on a variation of the Diff-VC \cite{diffvc} model. This approach leverages voice conversion (VC) techniques to enhance degraded speech signals. 
We use the pre-trained HuBERT \cite{hubert} encoder and vector quantization (VQ) with a cluster size of 2000 to extract content information from the input speech, instead of the phoneme-wise spectrogram averaging method in Diff-VC. This modification allows for more detailed and accurate content representation, which is crucial for high-quality speech restoration.

We utilize a transformer-based content encoder and a linear projection layer to convert the discrete HuBERT+VQ embeddings into a coarse spectrogram. The content encoder begins with three layers of convolution and layer normalization. Then, the output is passed through six layers of transformers, each with 192 channels and four attention heads. The output, projected into coarse spectrograms, serves as a content representation that bridges the gap between the degraded input and the restored output.

Next, using a U-Net decoder in the diffusion process, which takes the coarse spectrogram and speaker embedding, we obtain a restored spectrogram that closely matches the target. For the speaker encoder, we use the pre-trained ECAPA-TDNN \cite{ecapa} model, which provides robust and discriminative speaker representations. Finally, the pre-trained HiFi-GAN \cite{hifigan} is used to generate the speech waveform from the U-Net decoder output. 

The content encoder and U-Net decoder are trained using a weighted sum of two loss functions, as shown in Eq. (1). The encoder loss, $L_{enc}$, minimizes the L1-distance between $\bold{M}_{0}$ and $\bold{\hat{M}}$, the clean Mel-spectrogram, and the output of the content encoder after linear projection. This loss ensures that the content encoder accurately captures the essential features of the clean speech. The diffusion loss, $L_{d}$, is used to train the U-Net decoder, as shown in Eq. (3). This loss focuses on refining the spectrogram generated by the U-Net decoder, ensuring it closely matches the target spectrogram.

\begin{equation} L_{total} = L_{d} + \alpha L_{enc} \end{equation} \begin{equation} L_{enc}(\bold{M_0}, \hat{\bold{M}}) = \mathbb{D}(\bold{M_0}, \hat{\bold{M}}),
\end{equation} \begin{equation} L_d(\bold{M_{0}}) = \mathbb{E}{\epsilon_t}[|| s\theta(M_t, t | \hat{M}, \bold{s}) + \epsilon_{t}||^{2}_{2}]. \end{equation}

Here, $M_{t}$ and $\epsilon_{t}$ represent the Mel-spectrogram and noise at time step $t$, respectively. The score function, $s_\theta$, predicts the scale of the noise $\epsilon_{t}$, conditioned on the speaker embedding $\bold{s}$ and content embedding $\bold{\hat{M}}$. Classifier-free guidance \cite{guidance} is used during training with a probability of $p=0.1$ for speaker and content embeddings. This technique helps improve the robustness and generalization of the model. The parameter $\alpha$ is set to 0.5, balancing the contributions of the encoder and diffusion losses.

\begin{figure}
\includegraphics[width=\linewidth]{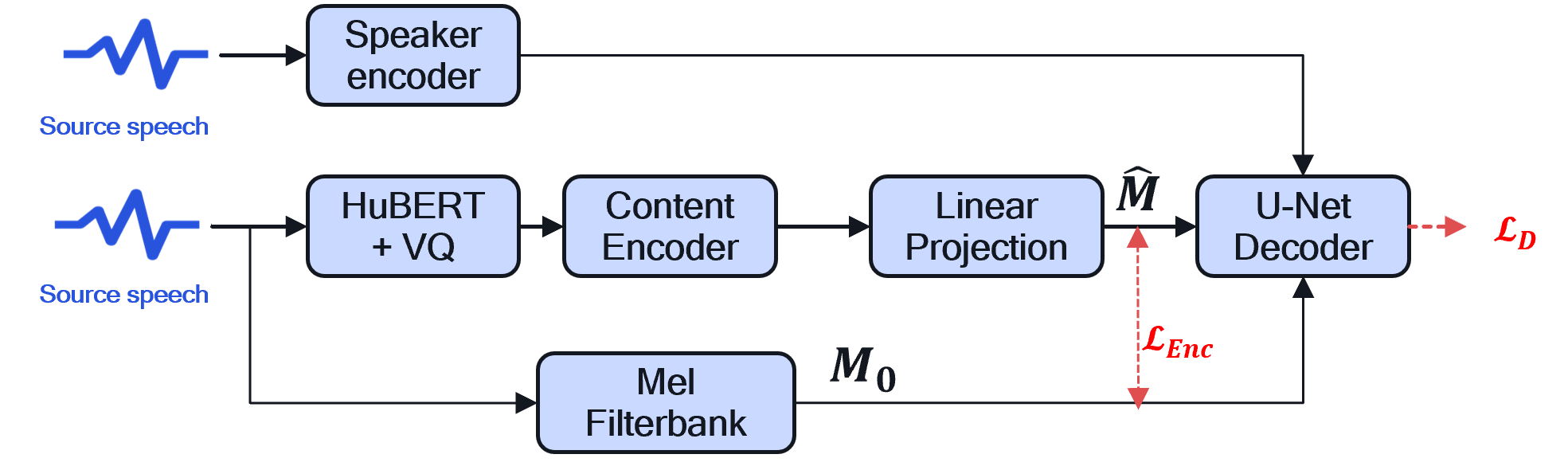}
\centering
\caption{\label{second}Training and content encoder adaptation framework for voice conversion-based speech restoration model.} 
\vspace{-15pt}
\end{figure}

\begin{figure}[t]
\begin{subfigure}{\columnwidth}
\centering
\includegraphics[width=0.85\textwidth]{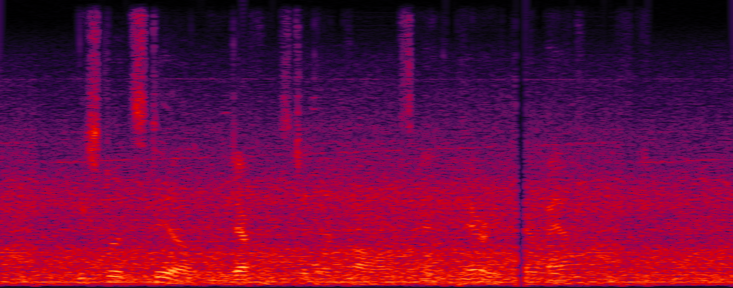}
\caption{\label{first} Noisy input spectrogram.} 
\end{subfigure}
\begin{subfigure}{\columnwidth}
\centering
\includegraphics[width=0.85\textwidth]{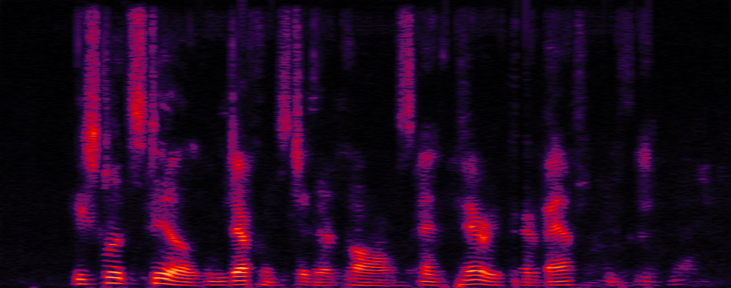}
\caption{\label{second} GSR model output.}
\centering
\end{subfigure}
\begin{subfigure}{\columnwidth}
\centering
\includegraphics[width=0.85\textwidth]{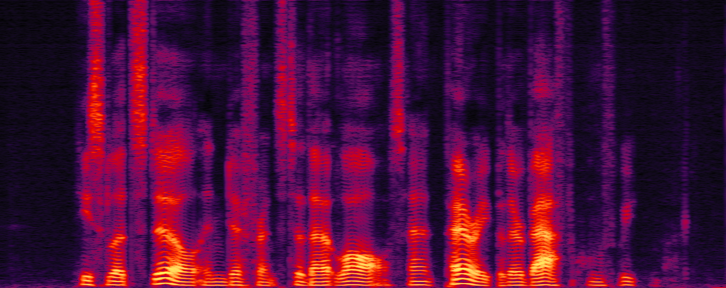}
\centering
\caption{\label{third} Proposed (GSR+VC) model output.}
\end{subfigure}
\vspace{-15pt}
\caption{Spectrogram comparison: (a) input mixture with noise and packet loss, (b) GSR output, and (c) final output from the proposed GSR+VC model.}
\label{overall}
\vspace{-15pt}
\end{figure}

\vspace{-5pt}
\section{Evaluation}
\subsection{Database and Training Setting}
GSR is trained on a proprietary speech enhancement dataset for 100 epochs with a batch size of 32, the AdamW \cite{loshchilov2017decoupled} optimizer and an initial learning rate of 2e-4, which decays exponentially with $\gamma = 0.999$. Each epoch contains 175 hours of noisy speech in the SNR range [-15, 40] dB. We used similar augmentations to those in \cite{liu2022voicefixer}, but also added codec artifacts \cite{koizumi2023miipher} and simulated packet-drop by randomly removing segments of 0-100 ms length in the time domain.
The VC model is trained on LibriTTS \cite{libritts}, which consists of 585 hours of clean speech from 2,456 speakers. We used the train-clean-100h and 360h subsets for training. The model is trained for 100 epochs with a batch size of 16 and the Adam-optimizer \cite{diederik2014adam} with a learning rate of 1e-4. 
For inference with the VC-inspired restoration model, we used scaling factors of 0.25 and 1.0 for the speaker and content embeddings, respectively, in classifier-free guidance, and performed 30 diffusion steps. The Mel-spectrogram parameters used in our experiments include a sample rate of 16,000 Hz, a window length of 1024, a hop length of 256, an FFT size of 1024, and 80 Mel bands.

We conducted experiments to verify the performance of the proposed method and compared it to three other state-of-the-art (SOTA) models: VoiceFixer (used as a baseline for GSR), FINALLY, and UNIVERSE (used for comparison with our final proposed system). We used two validation datasets for evaluation: VCTK-DEMAND \cite{valentini2017noisy} and UNIVERSE \cite{serra2022universal}.
The VCTK-DEMAND validation dataset contains one male and one female speaker, with additive noise from the DEMAND dataset mixed at 2.5, 7.5, 12.5, and 17.5 dB SNR. The UNIVERSE validation dataset consists of 100 audio clips formed from clean utterances sampled from VCTK and Harvard sentences, combined with noise and background sounds from DEMAND and FSDnoisy18k. This dataset includes various artificially simulated distortions, such as band limiting, reverberation, codec artifacts, and packet drops.
For evaluation, we employed NISQA (MOS pred), UTMOS, WV-MOS, and DNSMOS (OVRL) \cite{nisqa, saeki2022utmos, wv-mos, reddy2022dnsmos} as non-intrusive metrics to objectively assess the quality of the generated outputs.

\subsection{Comparison with Other Methods}
Miipher \cite{koizumi2023miipher} shares similarities to the proposed model also using speaker embeddings as guidance vectors when decoding waveforms. Miipher also employs a separate self-supervised model, PnG-BERT \cite{jia2021png}, to extract information from transcripts and clean up the w2vBERT \cite{chung2021w2v} features. Our approach differs in that we use a speaker-agnostic cleanup stage, which does not require transcript data or large self-supervised learning (SSL) models. This makes our method more efficient and less dependent on extensive computational resources.
UNIVERSE \cite{serra2022universal} is a diffusion-based speech restoration method. While the diffusion approach shows promise for high-quality speech restoration, it is not guided by speaker embeddings. Our GSR+VC method, on the other hand, leverages speaker-agnostic restoration combined with voice conversion, providing a robust solution that maintains high perceptual quality even in noisy conditions.
FINALLY \cite{babaev2024finally} utilizes GAN and WavLM-based SSL features and perceptual losses to improve the quality of speech recordings. The model, built upon the HiFi++ architecture, demonstrates state-of-the-art performance in producing clear, high-quality speech, effectively addressing various distortions such as background noise and reverberation. However, this model is also the largest in our comparison (454 M parameters).

\subsection{Discussion and Ablation Study} 
The overall performance of our proposed model (GSR+VC) is shown in Tables \ref{table:table1} and \ref{table:table2}. Additionally, we compared three systems in our ablation study (Table \ref{table:table3}): namely, GSR in standalone mode, VC in standalone mode (using both mel-spectrogram and SSL features as inputs), and combined GSR+VC, which is our proposed system. We observe that GSR on its own outperforms VoiceFixer on VCTK-DEMAND and on all four metrics on UNIVERSE, while being significantly smaller in size (Table \ref{table:table4}). This can be attributed to more training data, as well as the different neural vocoder used (HiFi-GAN for GSR and TFGAN \cite{tian2020tfgan} for VoiceFixer). GSR also considers more augmentations than VoiceFixer, e.g., codec artifacts and packet drops.

The experimental results on the VCTK-DEMAND and UNIVERSE validation sets demonstrate the effectiveness of our proposed GSR+VC method. On the VCTK-DEMAND set, our method achieves high scores across all metrics, with NISQA (4.24), UTMOS (4.12), WV-MOS (4.34), and DNSMOS (3.31). In the UNIVERSE validation set, the proposed method also achieves high scores on all metrics: NISQA (4.21), UTMOS (4.10), WV-MOS (4.23), and DNSMOS (3.26). It outperforms the baseline VoiceFixer and the clean speech reference. 
The combined GSR+VC system, outperforms UNIVERSE in 3 out of 4 metrics (Table \ref{table:table2}) while maintaining a similar model size (209 M vs 189 M parameters) and remains competitive with FINALLY, beating it in 2 out of 4 metrics on UNIVERSE (Table \ref{table:table2}) and exhibiting a better DNSMOS score on VCTK-DEMAND (Table \ref{table:table1}) whilst being less than half the size.
Finally, it is noted that despite the similarities between our proposed method and Miipher, conducting comparative experiments was challenging due to the unavailability of Miipher's data and model.

Fig. 3 illustrates an input example and stage-wise outputs of the proposed system.
When comparing Fig. 3 (a) and (b), the GSR model effectively reduces noise, however, it also introduces unwanted artifacts and distortions, that degrade the quality of the speech. Fig. 3 (c) shows how applying VC after the GSR output helps to further refine and restore the speech signal. The combined approach not only reduces noise but also successfully restores many of the damaged speech components seen in (b), leading to improved and more intelligible speech.

\begin{table}[!t]
    \caption{Experimental results on VCTK-DEMAND.}
    \vspace{-5pt}
    \resizebox{\columnwidth}{!}{%
    \centering
    \begin{tabular}{ccccc}
        \toprule
        \textbf{}  & \textbf{NISQA} & \textbf{UTMOS} & \textbf{WV-MOS} & \textbf{DNSMOS} \\
        \midrule
        \textbf{Input}         & 3.14 & 3.06 & 2.99 & 2.53 \\
        \textbf{Ground Truth}  & 4.09 & 4.07 & 4.50 & 3.15 \\
        \midrule
        \textbf{VoiceFixer \cite{liu2022voicefixer}}      & 3.81 & 3.71 & 4.18 & 3.13 \\ 
        \textbf{FINALLY \cite{babaev2024finally}}         & \textbf{4.48} & \textbf{4.32} & \textbf{4.87} & 3.22 \\
        \textbf{Proposed}       & 4.24 & 4.12 & 4.34 & \textbf{3.31} \\    
        \bottomrule
    \end{tabular}    
    }
    \vspace{-15pt}
    \label{table:table1}        
\end{table}

\begin{table}[!t]
    \caption{Experimental results on UNIVERSE.}
    \vspace{-5pt}
    \resizebox{\columnwidth}{!}{%
    \centering
    \begin{tabular}{ccccc}
        \toprule
        \textbf{} & \textbf{NISQA} & \textbf{UTMOS} & \textbf{WV-MOS} & \textbf{DNSMOS} \\
        \midrule
        \textbf{Input}                                  & 2.32 & 2.27 & 1.72 & 2.25 \\
        \textbf{Ground Truth}                           & 4.47 & 4.26 & 4.28 & 3.33 \\
        \midrule
        \textbf{VoiceFixer \cite{liu2022voicefixer}}     & 3.61 & 2.93 & 3.28 & 3.03 \\ 
        \textbf{UNIVERSE \cite{serra2022universal}}     & \textbf{4.30} & 3.89 & 3.85 & 3.22 \\
        \textbf{FINALLY \cite{babaev2024finally}}       & 4.20 & \textbf{4.19} & \textbf{4.40} & 3.23 \\
        \textbf{Proposed}                              & 4.21 & 4.10 & 4.23 & \textbf{3.26} \\ 
        \bottomrule
    \end{tabular}    
    }
    \vspace{-7.5pt}
    \label{table:table2}
\end{table}

\begin{table}[!t]
    \caption{Ablation study.}
    \vspace{-5pt}
    \resizebox{\columnwidth}{!}{%
    \centering
    \begin{tabular}{ccccc}
        \toprule
        \textbf{} & \textbf{NISQA} & \textbf{UTMOS} & \textbf{WV-MOS} & \textbf{DNSMOS} \\
        \midrule
        VCTK-DEMAND &  &  &  &  \\
        \midrule
        \textbf{GSR}        & 3.84 & 3.75 & 4.12 & 3.18 \\ 
        \textbf{VC (Mel)}   & 4.02 & 3.68 & 4.04 & 3.21 \\ 
        \textbf{VC (SSL)}   & 4.09 & 3.98 & 4.20 & 3.23 \\         
        \textbf{GSR+VC}     & \textbf{4.24} & \textbf{4.12} & \textbf{4.34} & \textbf{3.31} \\            
        \midrule
        UNIVERSE &  &  &  &  \\
        \midrule
        \textbf{GSR}        & 3.64 & 3.30 & 3.67 & 3.20 \\  
        \textbf{VC (Mel)}  & 3.33 & 2.83 & 3.16 & 2.93 \\ 
        \textbf{VC (SSL)}   & 3.96 & 3.72 & 4.04 & 3.15 \\         
        \textbf{GSR+VC}     & \textbf{4.21} & \textbf{4.10} & \textbf{4.23} & \textbf{3.26} \\            
        \bottomrule
    \end{tabular}
    }
    \vspace{-7.5pt}
    \label{table:table3}    
\end{table}

\begin{table}[!t]
    \caption{Comparison of model size and training data.}
    \vspace{-5pt}
    \resizebox{\columnwidth}{!}{%
    \centering
    \begin{tabular}{ccc}
        \toprule
        Model & Training Data & Model Size (parameters) \\
        \midrule
        \textbf{VoiceFixer \cite{liu2022voicefixer}}     & 44 hours (VCTK) & 117 M \\
        \textbf{GSR}     & 1750 hours (private data) & 40 M \\
        \textbf{UNIVERSE \cite{serra2022universal}} & 1500 hours (private data) & 189 M \\
        \textbf{FINALLY \cite{babaev2024finally}} & 200 hours (LibriTTS-R and DAPS) & 454 M \\
        \textbf{VC} & 585 hours (LibriTTS) & 169 M \\

        \bottomrule
    \end{tabular}
    }
    \vspace{-15pt}
    \label{table:table4}
\end{table}


For the ablation study, as can be seen in Table \ref{table:table3}, VC on its own is competitive, outperforming GSR across the board; however, by using GSR as a first stage, all metrics exhibit a substantial jump in quality for both datasets. To investigate the effectiveness of the SSL feature, we also tested VC (Mel), which directly uses the Mel-spectrogram instead of passing through the content encoder. In the VCTK-DEMAND dataset, the improvement was marginal. In the UNIVERSE dataset, VC (Mel) degraded performance. Given that the UNIVERSE dataset represents more severe conditions, the VC (Mel) system showed greater degradation under severe noise conditions, which indicates that the HuBERT+VQ process is contributing to the high perceptual quality score. 

While we have shown attractive speech restoration capabilities of our scheme, improving speech recognition system performance, for example, is another potential application of this approach. To this end, we plan to jointly train GSR, HuBERT, and the VC system in future studies. Furthermore, as alluded to in our ablation study, there is a difference between using Mel-spectrogram or SSL features as inputs to the VC system. In the future, we would like to explore using both features together.

\vspace{-3pt}
\section{Conclusion}
In this paper, we proposed an integrated framework that combines speech restoration and diffusion-based voice conversion (VC) to enhance speech quality. Our method leverages a two-stage approach, incorporating a generative speech restoration (GSR) model at the front end, followed by a VC module. This design addresses the vulnerability of VC models in noisy conditions and ensures high perceptual quality. The experimental results on the VCTK-DEMAND and UNIVERSE validation sets demonstrate the effectiveness of our proposed method, achieving high scores across multiple metrics. Our approach not only excels in perceptual quality and noise suppression, but also offers significant advantages in terms of efficiency and reduced dependency on transcript data, making it more practical for real-world applications.

\newpage


\bibliographystyle{IEEEtran}
\bibliography{main}
\end{document}